%% file: HD-7.tex
\documentclass[aps,superscriptaddress,reprint]{revtex4-1}
\usepackage[permil]{overpic}
\usepackage{graphicx}
\usepackage{latexsym}
\usepackage{amsmath}
\usepackage{amssymb}
\usepackage{upgreek}
\usepackage{xcolor}
\usepackage[colorlinks]{hyperref}
\hypersetup{allcolors=blue}
\begin{document}
\title{Extended defects in hard disk system and melting criteria.}
\author{M. V. Kondrin}
\email{mkondrin@hppi.troitsk.ru}
\affiliation{Institute for High Pressure Physics RAS, 108840 Troitsk, Moscow, Russia}
\author{Y.B. Lebed}
\affiliation{Institute for Nuclear Research RAS, 117312 Moscow, Russia}
\author{V. V. Brazhkin}
\affiliation{Institute for High Pressure Physics RAS, 108840 Troitsk, Moscow, Russia}

\begin{abstract}
The hard sphere model is widely used in description of fluids and solid media as a zero approximation to real systems. Despite the uniqueness of the model,  few analytical results are known for it, both for the 2D and 3D  cases. In present research we have investigated melting of the hard disk system by considering accumulation  of extended defects of a certain type in the crystaline phase, and jamming of the disk packing. It results in formulation of melting criteria with lower and upper bounds on volume ratio at melting transition: $25/21 \le V/V_0 \le 5/4$. It was found that, in full agreement with the Berezinskii-Kosterlitz-Thouless-Halperin-Nelson-Young theory, the 2D crystal melts into anisotropic liquid. The second transition, which is the transition between anisotropic and isotropic liquid has volume ratio $5/4 \le V/V_0 \le 13/9$.
\end{abstract}
\maketitle
Hard sphere model, beside being of academic interest and source of various mathematical problems, is successfully used in modeling of liquids, glasses, amorphous solids and granular materials, to name a few \cite{stillinger:s84,donev:jap04,torquato:rmp10}. The reason of such a versatility is that the hard sphere system is a good zero approximation in perturbation theory of disordered media \cite{kovalenko:pu73}. The idea is traced back to seminal paper of Zwanzig \cite{zwanzig:jcp54} and  was first successfully applied to liquids in the works of Barker and Henderson \cite{barker:jcp67,barker:jcp67a}. Still, despite an outstanding position of this model, there are few analytical results  for it, mainly obtained  through computer simulations \cite{lnp08}. This is true for the melting transition in the hard disk system (2D spheres) which has been modeled many times using Monte Carlo technique, since classical work of Alder and Wainwright \cite{alder:pr62}. In some sense 2D hard disk system is simpler, in others it is more complex  than its 3D counterpart. As it was suggested by the Nobel prize-winning Berezinskii-Kosterlitz-Thouless-Halperin-Nelson-Young theory, \cite{berez:jetp71,kosterlitz:jpc73,halperin:prl78,nelson:prb79,young:prb79}  the 2D isotropic liquid is expected to crystallize through a sequence of two continuous transitions -- the first (high-temperature) one is transition to anisotropic liquid (analog of liquid crystal), and the second one -- proper crystallization of anisotropic liquid into solid. Although, transition of anisotropic liquid into isotropic one is now believed to be first-order one \cite{bernard:prl14,qi:sm14}, the continuous character of the proper melting transition is never questioned.  This sequence is also observed in computer experiments \cite{mak:pre06,li:jcp22}.

However, simple criteria of melting in hard disk system are lacking. According to the Born criterion, the crystal melts when it can not longer  maintain shear stress. So, this brings into attention certain extended defects, which could form when the crystal is subjected to shear stress. Using geometrical consideration as well as energy of formation of such defects, one could speculate about the melting temperature. Such a strategy was applied  by us in a series of works where  we have estimated the melting temperature of crystals with diamond structure   \cite{kondrin:drm20,kondrin:pssb22,kondrin:jetp23} and graphene \cite{kondrin:prl21,kondrin:drm24}  using the density functional theory calculation. Our estimations mostly agree well with available experimental data. This shifts an attention from studying the liquids, which can be extremely difficult, to  investigation of certain type of crystal defects.

There is a subtle point here. The Born melting criterion gives the temperature of loss of stability of the crystal, i.e., the spinodal. For most crystals, no elastic moduli or phonon frequencies becomes  zero before melting. In other words, for melting, as for other transitions of the first order, a region of metastable states is possible on both sides of the transition, i.e. crystal overheating is possible, which is especially easy to observe in computer modeling, since the problem of the existence of the crystal surface is removed, from where melting mainly starts. It seems that for two-dimensional case, where the transition of infinite order into anisotropic liquid exists, melting and loss of stability generally coincide.

\begin{figure}
\includegraphics[width=\columnwidth]{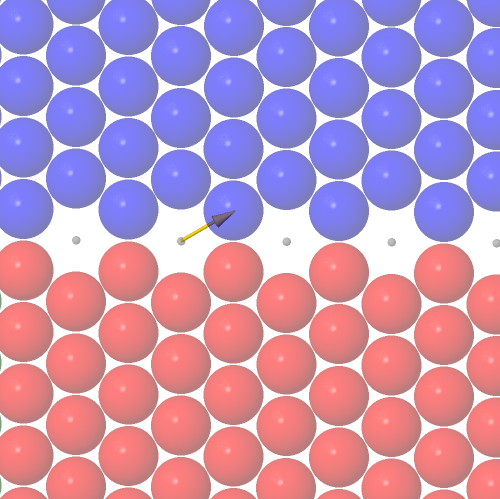}
\caption{Extended defect (crack) separating two half planes in otherwise intact triangular crystal lattice consisting of hard disks. Yellow arrow shows the shift of blue half plane from positions in the ideal crystal.}
\label{f:1}
\end{figure}

It is trivial to imagine the sort of extended defect which forms in triangular lattice (the crystal form of hard disk system) under shear stress. The sort of defects associated with stress relaxation is depicted in Fig.~\ref{f:1}. If we will assume that the disks have radius 1/2, such a defect (we will call it a crack) results in shifting one half plane along the other to the distance $\sqrt{3}/2$ along the crack and 1/2 normal to it. Using the popular notion in the close packing terminology, such a configuration is locally jammed (individual disks can not move) but not collectively jammed (two half planes can move with respect to each other). The less obvious result known from \cite{kondrin:drm24} is, what such defects can cross and so can form an irregular network of extended defects. The problem is what  the maximal concentration of defects in such a network is and what is the ratio $V/V_0$ where V is the volume (area) of defected crystal and $V_0$ that of ideal triangular lattice. Usually, when dealing with sphere packing the covering fraction $\eta$ and density $\rho$ are reported which are related to $V/V_0$ by the equations $V/V_0=\pi/(2\sqrt{3}\eta)=2/(\sqrt{3}\rho)$ valid for hard disk system. The anomaly observed on equation of state in computer simulations and associated with melting transition yields the value $V/V_0= 1.265-1.285$ \cite{li:jcp22}. In experiments on colloid solution (few micron-sized polymeric balls in water/alcohol mixture) in slit pore slightly different results are obtained $V/V_0=1.277-1.333$ \cite{thorneywork:prl17}. 

\begin{figure}
\includegraphics[width=\columnwidth]{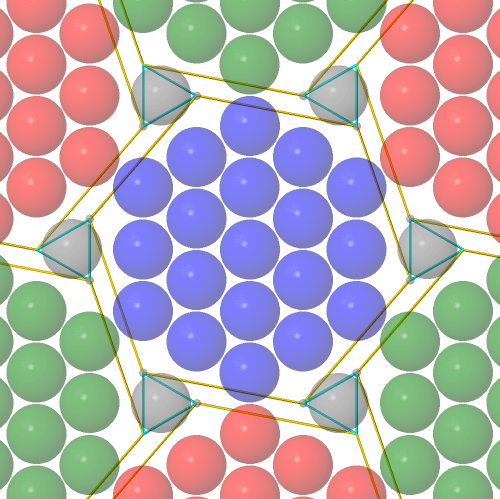}
\caption{Network of extended defects (cracks) in 19-core (shown in different colors) nearly jammed configuration. Grey balls stand for corner disks. Yellow and cyan lines are construction used for calculating $V/V_0$ ratio (see text). }
\label{f:2}
\end{figure}

The network of cracks with their maximal concentration is depicted in Fig.~\ref{f:2}. Proof that the concentration is maximal will be provided below. First we will focus on calculation the ratio $V/V_0$. $V_0$ is the area of yellow equilateral hexagon in Fig.~\ref{f:2} which is equal $21\sqrt{3}/2$. 21 is the number of disks contained in this area and it is the sum of disks strictly laying in the interior of hexagon (19 disks colored blue in Fig.~\ref{f:2} we will call it 19-core disks) and six disks in the corners (6 disks colored gray in Fig.~\ref{f:2} we will call it corner disks) which are shared between 3 hexagons meeting in the corner that is 2. The cracks area (that is $V-V_0$) is the area between hexagons and can be divided into corner unit triangles (drawn in cyan in Fig.~\ref{f:2}) with area $\sqrt{3}/4$ and the rest consisting of edge parallelograms with the area of each equal to $\sqrt{3}/2$. In calculating of $V$ the area of edge parallelogram should be divided by 2 and corner triangles by 3 taking into account the number of hexagons this area is shared by. This yields $V-V_0=2\sqrt{3}$ that is $V/V_0=25/21 \approx 1.190$.

Although as it can be concluded from the scrutiny of Fig.~\ref{f:2} 19-core disks retain contact with each other while corner disks can freely move inside triangle cages, this configuration is not jammed even locally because disks in the middle of 19-core edges can move but in a unique direction namely exactly normal to the core's edge. Still, in application to real systems in the sense of zero approximation of perturbation theory, we believe that attractive perturbing forces will glue together the 19-core disks, so during relaxation of shear stress the 19-core can move as a rigid body. We can call such a configuration nearly locally jammed. During the stress relaxation and formation of cracks 19-core disks retain their orientation. As  follows from comparison with experimental results and computer simulations , configuration with $V/V_0=25/21$ belongs to the crystal side of the melting transition. 

\begin{figure}
\includegraphics[width=\columnwidth]{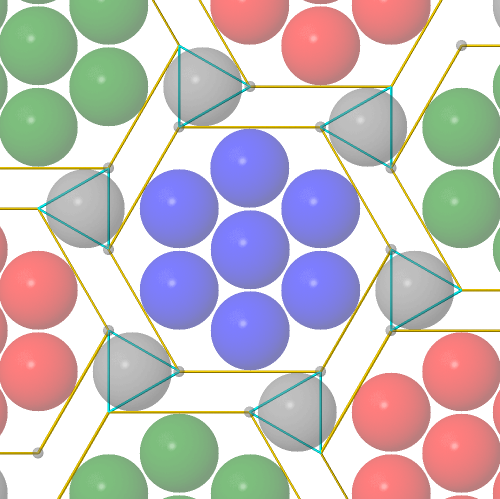}
\caption{Similar to Fig.~\ref{f:2} but with unjammed 7-core configuration.}
\label{f:3}
\end{figure}

What happens if we increase the concentration of cracks? More dense network of cracks with 7-core disks is depicted in Fig.~\ref{f:3}. Its ratio $V/V_0=13/9 \approx 1.444$ can be easily calculated taking into account, that cracks' area does not change in comparison to the previous case. However, on the other hand, from  Fig.~\ref{f:3} 7-core disks do not retain contacts with each other, so during  development of cracks the 7-cores regarded as rigid bodies can change their orientation. This corresponds to totally unjammed case. Therefore, this configuration is pertinent to description of isotropic liquid ,  which is corroborated by comparison with experiments and computer simulations.

It may seem that we are left with melting transition $V/V_0$ in the range $25/21 \le V/V_0 \le 13/9$. Still we were dealing with only isotropic liquid case, that is all cores were considered equilateral. If we remove this restriction, we will arrive to the elongated 14-core  case (by removing 5 disks from 2 adjacent edges of 19-core configuration). This configuration should be intermediate between 7-core and 19-core ones in the sense that some  disks  can be nearly jammed at their edges and some not. So we can conclude that the melting into anisotropic liquid with $25/21 \le V_m/V_0 \le 5/4$ and transition between isotropic and anisotropic liquids takes place at $5/4 \le V_i/V_0 \le 13/9$.

What does it correspond to in computer simulations? Anomaly observed on equation of state usually attributed to transition between isotropic and anisotropic liquids \cite{jaster:pla04,mak:pre06,bernard:prl14,li:jcp22}. Analysis of correlation of orientational order parameter yields the volume ratio value $V_i/V_0 \approx 1.290$ \cite{mak:pre06} which is close to lower bound of our transition criteria to isotropic liquid. Large margin of volume ratio $V_i/V_0$ to upper bound (our model overestimates stability region of anisotropic liquid) is not surprising taking into account the simplicity of the model based on considerations of extended defects in {\em crystal} phase. With proper melting transition the situation is worse because it does not manifest itself as an anomaly on the equation of state and requires for its determination correlation analysis of positional order parameter. Depending on the model assumed the different values of volume ratio is obtained: $V_m/V_0 \le 1.259$ \cite{bernard:prl14}, $V_m/V_0 \le 1.255$ \cite{mak:pre06} and $V_m/V_0 \approx 1.237$ \cite{jaster:pla04}, $V_m/V_0 \approx 1.250$ \cite{qi:sm14}. Thus, the first two values do not contradict us (we refine the upper bound) and the last two values nicely fit into the bounds of our melting criteria. 

To validate this model we try to estimate pressure in the limits of melting criteria. We start from combinatorial formula for Gibbs free energy due to permutations of $k$ defect layers of length $l$ similar to one depicted in Fig.~\ref{f:1} in the system of $N$ particles:

\begin{equation*}
\Delta F = P \Delta V - k_B T \log\left( \frac{n!}{k!(n-k)!}\right)
\end{equation*}

here $n=N/l$. Using the Stirling formula and substituting $\Delta V$ taking into account Fig.~\ref{f:1} we get:
\begin{equation*}
\begin{split}
\Delta F&=\frac{P k V_0}{2 n}- k_B T ( (n-k)\log(n)+k\log(n)-k\log(k) \\
&- (n-k) \log(n-k))
\end{split}
\end{equation*}

\begin{equation*}
\Delta F=\frac{P x V_0}{2} + k_B T n \left( (1 -x)\log(1-x)+x \log(x) \right)
\end{equation*}

here $x= k/ n$ -- defect concentration. We are interested in small defect concentrations, so we can reduce it to approximate equation:

\begin{equation*}
\Delta F \approx \frac{P x V_0}{2} + k_B T n x \log(x)
\end{equation*}

Before proceeding to minimization of this formula with respect to $x$ we modify it in application to general case with arbitrarily oriented extended defects (similar to one depicted in Fig.~\ref{f:2}). In this case $l$ (and therefore $n$) is not constant but depends on concentration $x$ by simple relation $l=\alpha/\sqrt{x}$. We should also include into the final formula for concentration certain configurational multiplier $Q=3$ which reflects the fact that in every corner three linear defects meet and the overall concentration of defects is $Q$ times larger than the one derived from the formula for defects aligned in single direction. So, the general formula has the form:

\begin{equation}
-\frac{PV_0}{3 k_B T N}= x^{1/2}/\alpha \log(x/Q) = \log(x/Q)/l
\label{eq:x}
\end{equation}

Substituting into this formula for $x=1-\rho$ the value $\rho \approx 0.923$ (corresponding to $V/V_0=5/4$) and $l=5/3$ (average side length of 14-core configuration) we get $\frac{PV_0}{k_B T N} \approx 6.61$. In broad sense this is in agreement with computer simulations which yield the value 8.3. This can be considered a good achievement for the simple model considering only one sort of defects. 

\begin{figure}
\includegraphics[width=\columnwidth]{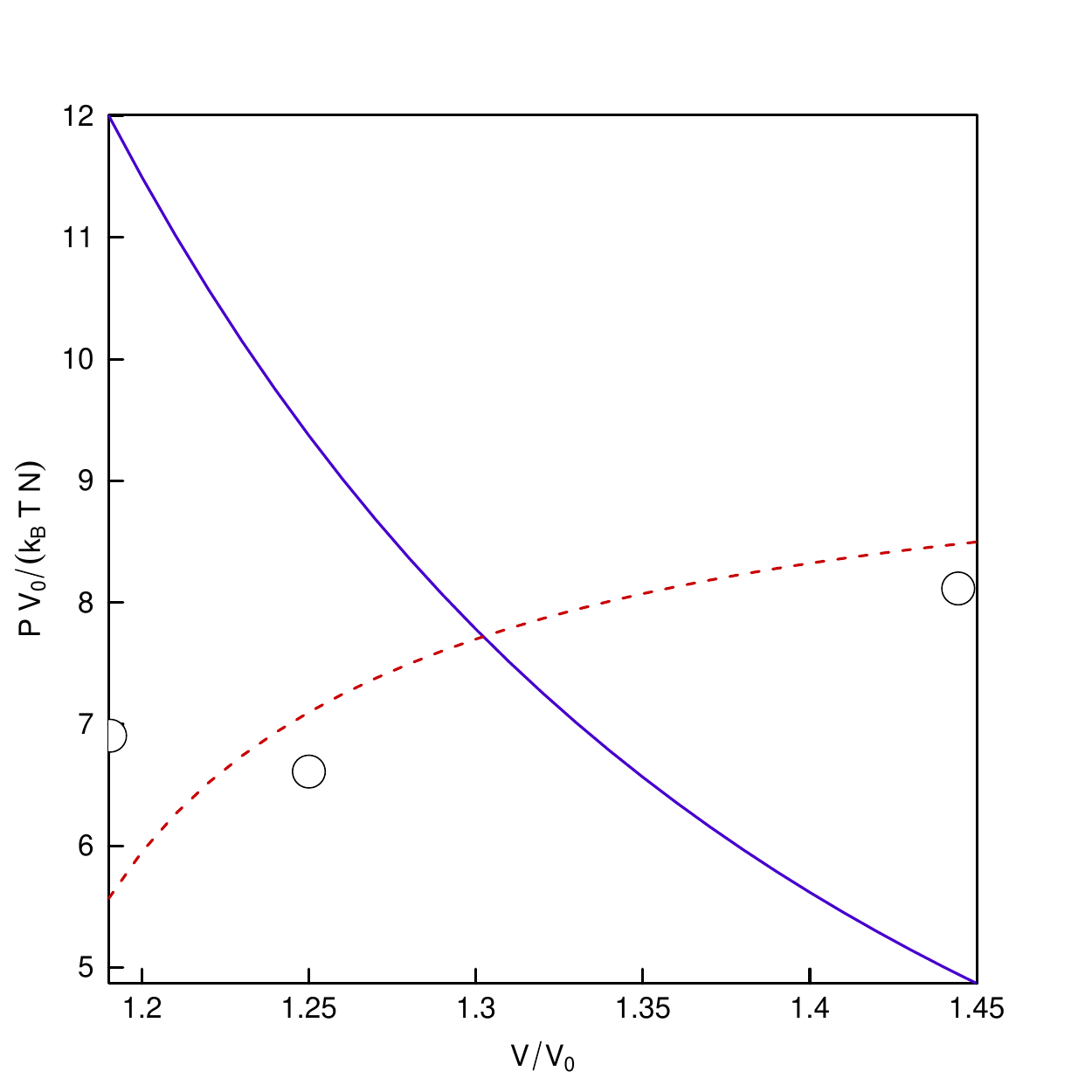}
\caption{Equation of state of hard disk system (solid blue curve) drawn according to virial expansion and compressibility factor according to Eq.~\ref{eq:x} in selected points (open circles). The symbols are fitted with smooth curve (dashed red) obtained by substitution $l=\alpha/\sqrt{x}$ in  Eq.~\ref{eq:x}.}
\label{f:4}
\end{figure}

We warn against taking Eq.~\ref{eq:x} as a formula for equation of state of hard disk system. Compelling reason for it is wrong asymptotic in $\rho \to 1$ limit: EOS in this limit is a diverging function but Eq.~\ref{eq:x} tends to zero. Rather Eq.~\ref{eq:x} should be regarded as another manifestation of melting criterion. Compressibility factor $\frac{PV_0}{k_B T N}$ according to Eq.~\ref{eq:x} is smaller in solid phase than predicted by EOS (say virial expansion \cite{lnp08}, see Fig.~\ref{f:4}) but the relation is inverted in liquid phase. This means that in solid phase the cracks are non equilibrium defects but always are in thermodynamic equilibrium in liquid phase. So, in liquid phase shear stress immediately produces cracks -- the liquid flows. The inversion point is located near $V/V_0 \approx 1.3$ that is near anomaly observed on equation of state in simulations.

To conclude, we demonstrated that accumulation of defects (cracks) in the hard disk system leads to different geometry of disks layout. At low defect concentrations the disks configuration is nearly jammed (and so can be attributed to the crystal phase) and at high concentration of defects configuration becomes totally unjammed (and thus pertinent to liquid). This leads us to the melting criteria with lower and upper bounds on volume ratio at melting transition $25/21 \le V_m/V_0 \le 5/4$. It was found that in full agreement with the Berezinskii-Kosterlitz-Thouless-Halperin-Nelson-Young theory the 2D crystals melts into anisotropic liquid. The second transition corresponding to transition between anisotropic and isotropic liquids has volume ratio $5/4 \le V_i/V_0 \le 13/9$.
%\bibliography{hd,2dlj,graphite}
\input{HD-7.bbl}
\end{document}